\documentclass[fleqn,usenatbib]{mnras}

\usepackage{newtxtext,newtxmath}
\usepackage[T1]{fontenc}

\usepackage{booktabs}
\usepackage{multirow} 
\usepackage{graphicx}

\usepackage{multicol}	
\usepackage[english]{babel}
\usepackage{epstopdf}
\usepackage{ae,aecompl}
\usepackage{hyperref}
\usepackage{amsmath}    % Advanced maths commands

\def\mpc{\,h^{-1}{\rm Mpc}}

\def\msun{\,h^{-1} M_\odot}

\usepackage{color}
\makeatletter

\newcommand{\Rmnum}[1]{\expandafter\@slowromancap\romannumeral #1@}
\makeatother

\title[mass bias in matched cosmic voids]{Multi-tracer mass bias in matched cosmic voids from SDSS DR7 and the ELUCID constrained simulation}

\author[Youcai Zhang et al.]{
Youcai Zhang$^{1}$\thanks{E-mail: yczhang@shao.ac.cn}
Xiaohu Yang$^{2,3}$\thanks{E-mail: xyang@sjtu.edu.cn}
Hong Guo$^{1}$,
and Peng Wang$^{1}$
\\
$^{1}${Shanghai Astronomical Observatory, Nandan Road 80, Shanghai 200030,
  China} \\
$^{2}$State Key Laboratory of Dark Matter Physics, Tsung-Dao Lee Institute \& School of Physics and Astronomy, Shanghai Jiao Tong University, \\~~~Shanghai 201210, China\\
$^{3}$Shanghai Key Laboratory for Particle Physics and Cosmology, and Key Laboratory for Particle Physics, Astrophysics and Cosmology, \\~~~Ministry of Education, Shanghai Jiao Tong University, Shanghai 200240, China 
}

\begin{document}
\label{firstpage}
\pagerange{\pageref{firstpage}--\pageref{lastpage}}
\maketitle

\begin{abstract}
Cosmic voids provide a unique environment for studying the relationship
between galaxies, subhaloes, and dark matter in the underdense Universe.
Using the SDSS galaxy catalogue and the ELUCID constrained simulation,
we establish an observationally anchored framework for measuring
multi-tracer mass bias within matched cosmic voids. A sample of 102
matched void pairs is constructed to compare galaxy, subhalo,
and dark matter mass distributions within an observationally constrained
realisation of the local Universe. 
We find that both the galaxy-to-dark matter and subhalo-to-dark matter
mass ratios decrease toward void centres, indicating that luminous and
halo tracers become increasingly depleted relative to the underlying
matter distribution in the deepest underdensities. In contrast, the
galaxy-to-subhalo mass ratio exhibits substantially larger statistical
uncertainties within the inner void regions
($r/R_{\rm v}\lesssim0.5$). By comparing measurements obtained using
independent and common coordinate frameworks, we show that coordinate
offsets contribute to the observed scatter but cannot fully account for
the large uncertainties. The remaining uncertainty primarily arises from
the severe scarcity of massive subhaloes
($\log_{10}(M_{\rm sub}/h^{-1}M_\odot)\ge11.8$) within void interiors,
which greatly reduces the number of statistically valid measurements near
void centres. Our results provide an empirical characterization of
multi-tracer mass bias in observationally constrained cosmic
environments and demonstrate the importance of accounting for
coordinate consistency and tracer scarcity when interpreting such
measurements in extreme underdense regions.

\end{abstract}


\section{Introduction}
\label{sec_intro}

Cosmic voids are the largest underdense structures in the Universe and occupy most of its volume \citep{Cautun2014, Contarini2026}. Owing to their low-density environments and relatively simple dynamical evolution, voids have become important laboratories for studying large-scale structure formation, galaxy evolution, environmental effects, and cosmological models \citep[e.g.][]{Sheth2004, Padilla2005, Hamaus2015, Hamaus2016, Argudo2026, Moretti2026a, Moretti2026b, Moretti2026c, Rouse2026, Takadera2026}. In recent years, increasing attention has been devoted to understanding how different tracers populate void environments and how their distributions relate to the underlying matter field \citep{Sutter2014, Nadathur2015b, Ghafour2025, Alfaro2026, London2026}.

Numerous studies have investigated the density profiles and
statistical properties of cosmic voids identified from galaxies,
dark matter haloes, and dark matter particles
\citep[e.g.][]{Colberg2005, Ricciardelli2014, Nadathur2015a,
Pollina2017, Pollina2019, Curtis2025, Song2026a}. These studies have demonstrated that inferred void
properties depend not only on the underlying matter distribution but
also on the tracer population used to define the void sample. Different tracers possess distinct bias and sampling characteristics,
which can lead to systematic changes in void size distributions and
radial density profiles, while leaving their three-dimensional
morphologies relatively stable \citep{Zhang2026}. Understanding these tracer-dependent effects is therefore essential
for interpreting observational void catalogues and for connecting
void statistics measured from luminous tracers to the underlying
matter distribution.

Despite these advances, comparisons between galaxies, haloes, and
dark matter within the same individual void environments remain
relatively limited. Observational studies are restricted to luminous
tracers and provide no direct access to the full matter field, whereas
numerical simulations offer complete information about the matter
distribution but do not generally reproduce the observed Universe on an
object-by-object basis. As a consequence, most previous comparisons have
been performed statistically, averaging over large void populations
rather than relating individual observed voids to corresponding systems
in simulations \citep{Hamaus2014, Sutter2015, Contarini2019, Ronconi2019, Tavasoli2026}. 
This limitation makes it difficult to assess how different
tracer populations populate the same underlying large-scale environment
and to quantify their relative mass contributions on a void-by-void
basis.

The ELUCID (\textit{Exploring the Local Universe with the reConstructed
Initial Density field}) project provides a particularly suitable framework for addressing this limitation.
ELUCID is a dark-matter-only constrained
simulation that reconstructs the large-scale matter distribution of the
local Universe using observational constraints derived from the Sloan
Digital Sky Survey (SDSS), generating a cosmological realisation that
closely reproduces the observed local cosmic web
\citep{Yang2007, Yang2012, WangHuiyuan2012, WangHuiyuan2014,
WangHuiyuan2016, WangHuiyuan2018}. Although the exact positions of individual galaxies and subhaloes
are not expected to be reproduced on small scales, their reconstructed large-scale environments correspond closely
to those observed within the SDSS volume
\citep{Yang2018, Zhang2021a, Zhang2021b, Zhang2022, Zhang2024, Zhang2025, 
Zhang2026}. This correspondence provides a basis for associating observed
voids with counterpart voids in the constrained simulation while retaining
access to the underlying dark matter distribution and halo populations
that are unavailable observationally.

The matched-void framework therefore provides a complementary approach
to conventional multi-tracer void analyses, in which void catalogues
identified from different tracer populations are compared only at the
population level. By associating an observed galaxy-defined void with a
counterpart in an observationally constrained simulation, the framework
allows the galaxy, subhalo, and dark matter distributions to be examined
within corresponding large-scale environments. This preserves the
observational connection of the galaxy sample while providing access to
the halo and dark matter information available in the simulation. It also
makes it possible to quantify the relative mass contributions of different
tracers and to assess how differences in void geometry and tracer sampling
affect such comparisons. 

In this work, we combine a volume-limited SDSS galaxy sample with the
ELUCID constrained simulation to investigate the relative mass content
of galaxies, subhaloes, and dark matter in matched cosmic voids. Voids
are independently identified from the observed galaxy distribution and
the ELUCID subhalo catalogue, and corresponding systems are associated
through a one-to-one cross-matching procedure. Using the resulting
matched-void sample, we measure three radial mass-ratio statistics,
$\mathcal{R}_{\rm g/dm}$, $\mathcal{R}_{\rm sub/dm}$, and
$\mathcal{R}_{\rm g/sub}$, which quantify the radial variation of the relative mass contributions
of different tracer populations within the same matched void
environments.

To evaluate the impact of geometric differences between independently
identified void catalogues, we perform the measurements under both an
independent-frame scheme, which preserves the native void configurations
of the SDSS and ELUCID catalogues, and a common-frame scheme, in which
all tracer populations are measured within a shared geometric framework.
This allows us to investigate how differences in void centres and
effective radii between the independently identified SDSS and ELUCID
void counterparts affect the inferred mass-ratio measurements, and to
examine the statistical limitations associated with the scarcity of
tracer populations within the deepest void interiors.

Throughout this paper, we adopt the cosmological parameters used in the
ELUCID simulation: $\Omega_m = 0.258$, $\Omega_b = 0.044$,
$\Omega_\Lambda = 0.742$, $h = 0.72$, $n_s = 0.963$, and
$\sigma_8 = 0.796$. All quantities used in the SDSS and ELUCID analyses,
including the stellar mass estimates, are consistently expressed under
this cosmology.

This paper is organized as follows. Section~\ref{sec_data} describes the
SDSS galaxy sample and the ELUCID constrained simulation.
Section~\ref{sec:methods} presents the void identification procedure,
the construction of the matched-void catalogue, and the methodology
used to measure multi-tracer mass-ratio profiles.
Section~\ref{sec:results} presents the properties of the matched void
sample and the resulting mass-ratio measurements.
Section~\ref{sec:discussion} discusses the implications of these
results for multi-tracer void studies.
Finally, Section~\ref{sec:conclusion} summarizes the main conclusions
of this work.

\section{Observational Data and Constrained Simulation}
\label{sec_data}

\subsection{Galaxies from SDSS DR7}
\label{obs_data}

The observational galaxy sample used in this work is drawn from the New York University Value-Added Galaxy Catalogue \citep[NYU-VAGC;][]{Blanton2005}, based on the Sloan Digital Sky Survey Data Release 7 \citep[SDSS DR7;][]{York2000,Abazajian2009}.

Following \citet{Shi2016}, we restrict the analysis to the contiguous region in the Northern Galactic Cap (NGC), covering
$111^{\circ}<\alpha<264^{\circ}$ and
$-3^{\circ}<\delta<68^{\circ}$.
A volume-limited sample is constructed within the redshift range
$0.01 \le z \le 0.12$
and with an $r$-band absolute magnitude limit
$^{0.1}M_r - 5\log h < -20.09$,
where magnitudes are $K$-corrected and evolution-corrected to $z=0.1$ \citep{Blanton2003a,Blanton2003b}. Galaxy stellar masses are estimated using the colour-dependent mass-to-light ratio relation of \citet{Bell2003}.

To identify voids in real space, we adopt the reconstructed galaxy catalogue of \citet{Shi2016}, in which redshift-space distortions (RSD) are approximately corrected using a group-based reconstruction method. Detailed descriptions of the reconstruction procedure are provided in \citet{Shi2016} and are not repeated here. The resulting real-space catalogue contains 158,840 galaxies and is identical to the sample analysed in \citet{Zhang2026}.

\subsection{Dark matter subhaloes from ELUCID}
\label{sec:elucid_sim}

To provide a dynamically reconstructed matter distribution for comparison
with the observed galaxy sample, we utilize the ELUCID constrained
simulation \citep{WangHuiyuan2014, WangHuiyuan2016}. ELUCID is designed to
reproduce the large-scale structure of the local Universe by incorporating
observational constraints derived from SDSS galaxy groups. The resulting
constrained realisation enables a direct comparison between observed
galaxies, simulated subhaloes, and the underlying dark matter distribution
within a common cosmological volume.

The ELUCID reconstruction combines approximate RSD
correction, halo-domain density reconstruction, Hamiltonian Monte Carlo
(HMC) optimization, and subsequent $N$-body evolution
\citep{WangHuiyuan2009, WangHuiyuan2013}. The simulation
was performed in a periodic box of $500\,h^{-1}{\rm Mpc}$ on a side using
$3072^3$ dark matter particles, with initial conditions generated at
$z=100$. The subsequent evolution to the present epoch was carried out
using a memory-optimized version of \texttt{GADGET-2}
\citep{Springel2005}. The resulting particle mass resolution is
$3.1\times10^8 \msun$. Dark matter haloes and subhaloes were
identified using the Friends-of-Friends (FOF) algorithm
\citep{Davis1985} and the \texttt{SUBFIND} algorithm
\citep{Springel2001}.

The observational constraints used in the ELUCID reconstruction are
primarily derived from SDSS galaxy groups with halo masses above
$10^{12}\,h^{-1}M_{\odot}$ \citep{Yang2007, Yang2012}. Therefore, the
reconstruction is expected to be most reliable on large and intermediate
scales, where the cosmic web is strongly constrained by the observed
environment. Previous studies have demonstrated that ELUCID successfully
reproduces prominent nearby structures, including the Coma cluster and the
Sloan Great Wall \citep{WangHuiyuan2016, Chen2019, Luo2024}.

In addition to the dark matter particle distribution, we construct a
subhalo catalogue with a number density matched to that of the SDSS galaxy
sample. Specifically, we impose a lower mass threshold of
$10^{11.8}\msun$ for individual subhaloes, which yields a
subhalo abundance consistent with the volume-limited SDSS galaxy
catalogue described in Section~\ref{obs_data}. The resulting
subhalo sample contains the same total number of objects as the observed
galaxy sample within the reconstructed survey volume. This selection
minimizes differences caused purely by tracer sampling density and allows
a more direct comparison between galaxy and subhalo populations.

Although the large-scale density field of ELUCID is constrained by
observations, the exact positions of individual galaxies and subhaloes are
not expected to coincide on small scales. The subhalo population results
from nonlinear gravitational evolution within the reconstructed density
field and therefore represents a statistically consistent realisation of
the local Universe rather than an object-by-object reproduction of the
observed galaxy distribution. Consequently, voids identified from galaxies
and subhaloes may exhibit differences in their centres, effective radii,
and detailed boundaries even when they trace the same underlying
large-scale underdensity. Quantifying these geometric differences motivates
the matched-void analysis presented in this work.

The combination of observed galaxies, abundance-matched subhaloes, and the
underlying dark matter distribution within the same constrained volume
provides a framework for investigating multi-tracer mass bias in cosmic
voids.

\section{Void Analysis and Statistical Methods}
\label{sec:methods}

\subsection{Void identification}
\label{subsec:voids}

Cosmic voids are identified using the
\texttt{REVOLVER} pruning mode \citep{Nadathur2019}
implemented within the \texttt{VAST} framework
\citep{Douglass2023}, which identifies watershed-based underdense regions from a Voronoi
tessellation of the tracer distribution.

Each void is characterised by its volume-weighted centre,
\begin{equation}
\label{eqn:center}
\mathbf{C} =
\frac{\sum_i \mathbf{x}_i V_i}
     {\sum_i V_i},
\end{equation}
where $\mathbf{x}_i$ and $V_i$ are the position and Voronoi cell volume of the $i$-th tracer, respectively. The effective void radius is defined as
\begin{equation}
R_{\rm v} =
\left(
\frac{3V_{\rm void}}
     {4\pi}
\right)^{1/3},
\end{equation}
with $V_{\rm void}$ denoting the total watershed volume associated with each void.

Voids are identified independently in two datasets: the SDSS galaxy
distribution in real space and the ELUCID reconstructed subhalo field.
Although both tracer populations trace the same underlying large-scale
structure and are selected to have comparable number densities,
differences in tracer bias, sampling effects, and reconstruction
uncertainties can still lead to variations in void membership and
geometry.

In this work, we adopt the \texttt{REVOLVER} void definition.
Previous analyses by \citet{Zhang2026} showed that, among the void
definitions implemented in the \texttt{VAST} framework,
\texttt{REVOLVER} yields more consistent stacked radial number-density
profiles between SDSS galaxies and ELUCID subhaloes than
\texttt{VoidFinder} \citep{Hoyle2002}. This result indicates that \texttt{REVOLVER} yields 
a higher level of consistency between the SDSS and ELUCID void populations, making it
particularly suitable for the comparative analysis performed in this
work.

Following \citet{Douglass2023}, we retain only voids with effective radii
\begin{equation}
R_{\rm v} \ge 10\,h^{-1}{\rm Mpc},
\end{equation}
for the scientific analysis. The initial void search is performed down to
$5\,h^{-1}{\rm Mpc}$ to improve completeness near the selection threshold.

To minimize contamination from survey and reconstruction boundaries, we
further restrict the sample to interior voids. Following the \texttt{VAST}
classification scheme, a candidate void is retained only if its
edge-to-total area ratio is smaller than 0.1. This criterion removes voids
whose watershed basins are significantly affected by the survey boundary and
ensures that the measured void properties are dominated by genuine
large-scale underdensities rather than edge artifacts.

After applying these selection criteria, the final samples consist of
389 voids in SDSS real space and 404 voids in the ELUCID reconstructed
volume. A subset of these systems is subsequently used to construct the
one-to-one matched catalogue employed throughout the remainder of this work.

\subsection{Void cross-matching}
\label{subsec:matching}

To establish a reliable counterpart link between the observational and simulated cosmic webs, we develop a one-to-one cross-matching pipeline to connect individual voids identified in the SDSS galaxy catalogue with those in the ELUCID simulation. 

We implement a multi-stage search across the two void populations, which proceeds as follows:
\begin{enumerate}
    \item \textit{Proximity Filtering}: For any given candidate pair consisting of an SDSS void and an ELUCID void, we first enforce a spatial proximity constraint. The three-dimensional distance $d=|\mathbf{C}_{\rm SDSS} - \mathbf{C}_{\rm ELUCID}|$ between their volume-weighted centres must satisfy
    \begin{equation}
    d < 0.5 \times \min(R_{\rm SDSS}, R_{\rm ELUCID}),
    \end{equation}
    where $\mathbf{C}_{\rm SDSS}$ and $R_{\rm SDSS}$ denote the volume-weighted centre and effective radius of the considered SDSS void, while $\mathbf{C}_{\rm ELUCID}$ and $R_{\rm ELUCID}$ represent those of the candidate ELUCID void, with basic quantities defined following Section~\ref{subsec:voids}.

    \item \textit{Volumetric Overlap Quantification}: For all candidate pairs surviving the proximity filter, we evaluate their geometric similarity using the Intersection-over-Union (IoU) metric. To obtain a computationally efficient estimate of the overlap, the irregular watershed voids are approximated as spheres with radii equal to their effective radii. Since the IoU
    criterion is used only as an initial matching filter rather than as a precise geometric measurement, this approximation is sufficient for identifying candidate counterpart systems.
    The overlap volume, $V_{\rm intersect}$, is then computed analytically from the centre separation using the standard sphere--sphere intersection formula. The IoU is defined as
    \begin{equation}
    {\rm IoU} =
    \frac{V_{\rm intersect}}
    {V_{\rm SDSS}+V_{\rm ELUCID}-V_{\rm intersect}},
    \end{equation}
    where $V_{\rm SDSS}$ and $V_{\rm ELUCID}$ are the volumes of the corresponding equivalent spheres defined by $R_{\rm SDSS}$ and $R_{\rm ELUCID}$, respectively. 
    We retain only candidate pairs with ${\rm IoU}>0.4$, ensuring substantial volumetric overlap in addition to the proximity criterion.  We tested the
    sensitivity of the matching procedure to the adopted IoU threshold by repeating the analysis with ${\rm IoU}>0.3$ and ${\rm IoU}>0.5$. The resulting mass-ratio measurements show consistent radial trends with those obtained for the fiducial threshold of 0.4, indicating that the main conclusions are insensitive to reasonable variations in the IoU threshold.

    \item \textit{Mutual Best-Match Verification}: 
    To resolve potential multi-matching ambiguities and establish a strictly bijective mapping, 
    we identify the highest-IoU counterpart for each SDSS void and, independently, the highest-IoU counterpart for each ELUCID void. A pair is retained only if the two voids mutually identify 
    each other as their preferred match.
    
\end{enumerate}

Applying the above procedure yields a final catalogue of mutually matched
void pairs. We verified that the resulting matched catalogue is insensitive to the
specific matching implementation. In particular, standard one-way
best-IoU matching, mutual best-match selection, and a size-sorted
greedy implementation all recover the same matched sample.
The statistical and geometric properties of the final matched catalogue
are presented in Section~\ref{subsec:geometric_diagnostics}.

The matching procedure is essential for the present analysis because
the objective is to measure the relative mass contributions of different
tracers within the same cosmic environments. Without matching, voids
identified from different tracer populations cannot be assumed to
represent identical underdense regions, because tracer bias and sampling
effects modify both the spatial distribution of tracers and the resulting
void geometry. In such cases, the measured mass ratios would combine the
intrinsic tracer--matter relation with differences in the underlying
environments sampled by the two tracer populations.

\begin{figure*}
\centering
\includegraphics[width=0.95\textwidth]{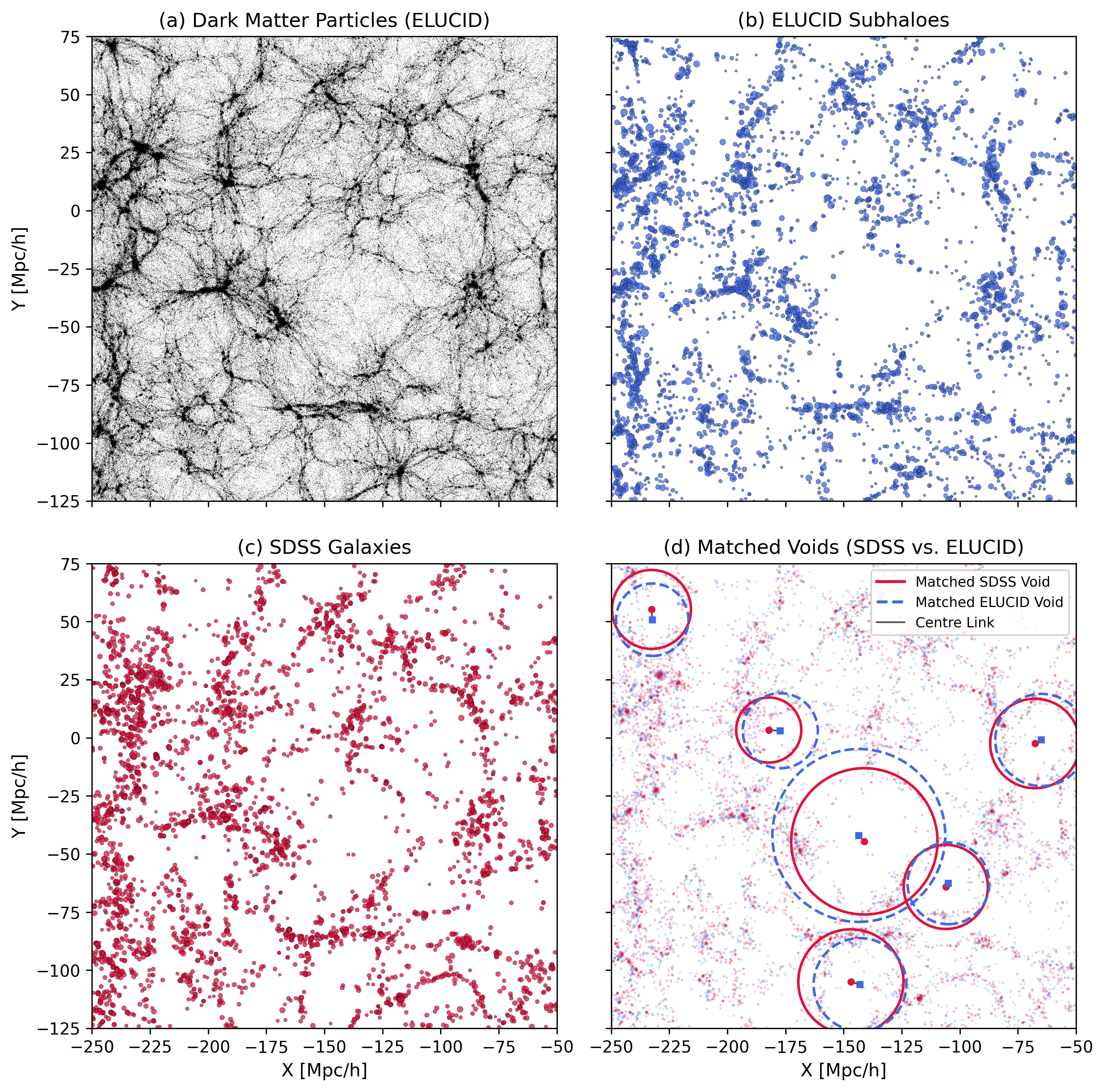}
\caption{
Comparison of the cosmic web and void structures between the ELUCID
constrained simulation and SDSS observations presented within a
$15\mpc$ thick spatial slice.
Panel (a): Distribution of dark matter particles ($1\%$ random sampling)
within the ELUCID simulation, delineating the underlying matter field of
the cosmic web.
Panel (b): Spatial distribution of ELUCID subhaloes selected above a
threshold of $10^{11.8}\msun$, chosen to match the number
density of the observed galaxy sample. The marker size scales
monotonically with the subhalo mass.
Panel (c): Spatial distribution of SDSS bright galaxies
($^{0.1}M_r - 5\log h < -20.09$), with marker sizes proportional to
their stellar masses.
Panel (d): Matched cosmic voids extracted using the
\texttt{REVOLVER} algorithm. Solid red and dashed blue circles denote the
effective spherical boundaries of the SDSS and ELUCID void populations,
respectively, while grey solid lines connect the centres of mutually
matched pairs. Note that overlapping circles result from the projection
of three-dimensional watershed-based structures.}
\label{fig:void_slice}
\end{figure*}

\subsection{Mass-ratio measurements}
\label{subsec:mass_ratio}

For each matched void pair, we measure the radial distributions of galaxy mass
($M_{\rm g}$), dark matter mass ($M_{\rm dm}$), and subhalo mass
($M_{\rm sub}$) within concentric spherical shells extending to
$3R_v$. All measurements are performed in bins of the normalised radius
$r/R_v$, where $R_v$ denotes the effective void radius.

To quantify the relative distributions of the three tracers, we define the
following mass-ratio statistics:
\begin{equation}
\mathcal{R}_{\rm g/dm}(r)
=\frac{M_{\rm g}(r)}
{M_{\rm dm}(r)},
\end{equation}
\begin{equation}
\mathcal{R}_{\rm sub/dm}(r)
=\frac{M_{\rm sub}(r)}
{M_{\rm dm}(r)},
\end{equation}
\begin{equation}
\mathcal{R}_{\rm g/sub}(r)
=\frac{M_{\rm g}(r)}
{M_{\rm sub}(r)}.
\end{equation}
The mean radial mass-ratio profile is estimated from stacked masses,
\begin{equation}
\label{eqn:stacked_mean}
\langle \mathcal{R} \rangle_j=\frac{\sum_{i=1}^{N_{\rm v}} A_{i,j}}
{\sum_{i=1}^{N_{\rm v}} B_{i,j}},
\end{equation}
where $A_{i,j}$ and $B_{i,j}$ represent the total integrated mass of the numerator and denominator 
tracers (i.e. galaxy, subhalo, or dark matter
mass, depending on the estimator) within the $j$-th radial
bin of the $i$-th void pair, and $N_{\rm v}$ is the total number
of matched voids.

This approach is adopted because the innermost regions of cosmic voids are
intrinsically underdense and may contain very small amounts of tracer mass.
Stacking masses before taking ratios provides a stable estimate of the global
mass-ratio profile and avoids numerical instabilities associated with empty or
nearly empty bins.

The uncertainty is quantified using the standard error of the mean (SEM)
derived from the distribution of individual void mass ratios,
\begin{equation}
{\rm SEM}(\mathcal{R})_j=\frac{\sigma(\mathcal{R}_{i,j})}
{\sqrt{N_j}},
\end{equation}
where $\mathcal{R}_{i,j}=A_{i,j}/B_{i,j}$ is the mass ratio measured for the
$i$-th void in the $j$-th radial bin, $\sigma(\mathcal{R}_{i,j})$ is the
corresponding sample standard deviation, and $N_j$ is the number of valid
voids contributing to that radial bin. Only voids with non-zero denominator
masses are included in the SEM calculation.

Consequently, the mass-ratio profiles presented in this work represent ratios of
stacked masses, while the error bars quantify the void-to-void scatter of the
corresponding individual mass ratios. The two statistics therefore probe
different aspects of the matched-void population: the stacked profiles trace
the mean mass distribution, whereas the SEM reflects the intrinsic diversity
among individual void systems.

\begin{figure*}
\centering
\includegraphics[width=\textwidth]{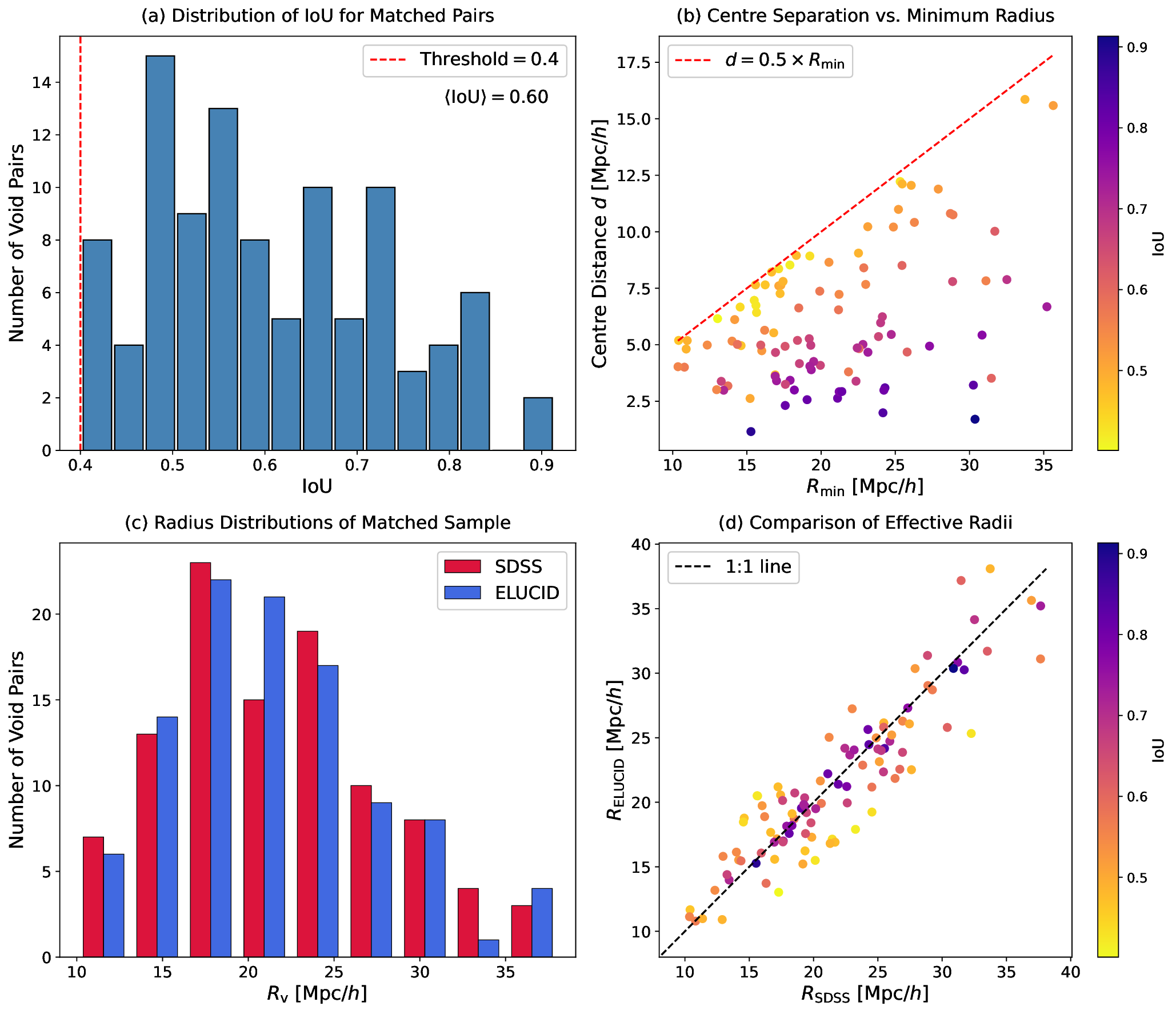}
\caption{
Statistical and geometric diagnostics of the 102 mutually matched
SDSS--ELUCID void pairs.
Panel (a): Distribution of the Intersection-over-Union (IoU) values,
where the red dashed line marks the adopted selection threshold
(${\rm IoU}>0.4$).
Panel (b): Three-dimensional centre separation as a function of the
minimum effective radius,
$R_{\rm min}\equiv\min(R_{\rm SDSS},R_{\rm ELUCID})$,
bounded by the proximity criterion (red dashed line). Data points are
colour-coded according to their individual IoU values.
Panel (c): Differential size distributions of the matched SDSS and
ELUCID void populations, showing good agreement between the two samples.
Panel (d): Direct comparison of the effective radii of matched
counterparts, where the grey dashed line represents the ideal
one-to-one correspondence. The effective radii are strongly correlated,
with a Pearson correlation coefficient of $0.91$.
}
\label{fig:void_matching_diagnostics}
\end{figure*}

\section{Results}
\label{sec:results}

\subsection{Visual inspection of matched voids}
\label{subsec:visual_inspection}

Figure~\ref{fig:void_slice} provides an overview of the tracer
populations and representative matched void systems used throughout
this work.

Panel (a) shows the dark matter distribution extracted from the ELUCID
constrained simulation, while Panels (b) and (c) display the
corresponding ELUCID subhalo and SDSS galaxy populations within the same
survey volume, respectively. Major filaments, clusters, and
underdense regions can be identified in all three tracer distributions,
indicating that the ELUCID reconstruction successfully reproduces the
principal large-scale structures observed in the local Universe.

Panel (d) presents several representative matched void pairs identified
using the procedure described in Section~\ref{subsec:matching}. Solid
red circles denote SDSS voids and dashed blue circles represent their
ELUCID counterparts, while grey lines connect the centre positions of
mutually matched pairs. Visual inspection shows that the matched systems
generally occupy the same large-scale underdense environments, supporting
the overall robustness of the matching procedure.

Nevertheless, noticeable differences remain in both the centre positions
and effective radii of individual matched voids. These differences are
expected because the two void catalogues are identified independently
from different tracer populations. The statistical properties of these
matched systems, including their volumetric overlap, centre offsets, and
radius correspondence, are quantified in the following subsection.

\subsection{Geometric diagnostics of matched voids}
\label{subsec:geometric_diagnostics}

To quantify the properties of the matched sample, Figure
~\ref{fig:void_matching_diagnostics} presents several diagnostic
statistics for the 102 matched SDSS--ELUCID void pairs.

Panel (a) shows the distribution of the IoU values among the matched
void pairs. By construction, all matched systems satisfy the adopted
threshold of ${\rm IoU}>0.4$, with a mean value of
$\langle{\rm IoU}\rangle=0.60$. The distribution indicates that most
matched pairs exhibit substantial volumetric overlap.

Panel (b) shows the three-dimensional centre separation as a function
of the minimum effective radius,
$R_{\rm min}=\min(R_{\rm SDSS},R_{\rm ELUCID})$.
All matched systems satisfy the proximity criterion
$d<0.5R_{\rm min}$,
although a broad range of centre offsets remains within this limit.
The colour coding indicates the corresponding IoU values, showing that
pairs with larger centre separations generally exhibit lower volumetric
overlap, whereas the highest-IoU matches tend to have smaller spatial
offsets.

The size distributions of the matched void populations are shown in
Panel (c). The SDSS and ELUCID samples exhibit very similar radius
distributions over the full matched range. This agreement is further
illustrated in Panel (d), which compares the effective radii of
individual matched pairs. The two measurements are strongly correlated,
with a Pearson correlation coefficient of
$0.91$.

Overall, the matched SDSS and ELUCID voids occupy similar large-scale
underdense regions and exhibit comparable characteristic sizes.
Nevertheless, non-negligible differences remain in their centre
positions and, to a lesser extent, their effective radii.
Because the mass-ratio measurements are defined relative to these
quantities, such geometric discrepancies may introduce additional
scatter into the inferred radial profiles, particularly within the
innermost void regions. To assess the importance of this effect, we
first present measurements performed in the native coordinate systems
of the two catalogues before introducing an alternative common-frame
analysis.

\subsection{Independent-frame measurements}
\label{subsec:independent_frame}

\begin{figure*}
\centering
\includegraphics[width=\textwidth]{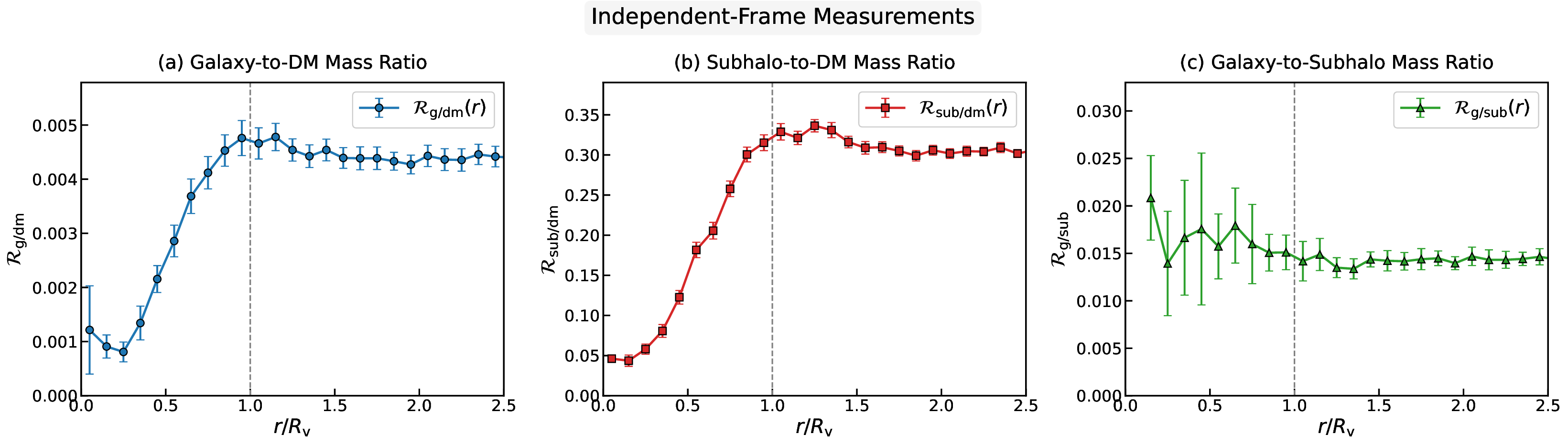}
\caption{
Radial profiles of the stacked mass ratios measured for the matched-void sample ($N_{\rm v}=102$) using the independent-frame scheme. Panels (a), (b), and (c) show the galaxy-to-dark matter ($\mathcal{R}_{\rm g/dm}$), subhalo-to-dark matter ($\mathcal{R}_{\rm sub/dm}$), and galaxy-to-subhalo ($\mathcal{R}_{\rm g/sub}$) mass ratios, respectively. The vertical dashed line indicates the effective void boundary ($r/R_{\rm v}=1$). Error bars denote the standard error of the mean (SEM) derived from the distribution of individual void ratios. Radial bins containing fewer than five valid void pairs are omitted.
}
\label{fig:void_mass_ratios}
\end{figure*}

We first investigate the radial mass-ratio profiles measured from the full
matched-void sample ($N_{\rm v}=102$) using the independent-frame scheme.
Under this configuration, galaxy masses are measured relative to the native
SDSS void centres and radii, whereas dark matter and subhalo masses are
evaluated relative to their corresponding ELUCID void definitions. This approach
preserves the native coordinate systems of the two independently constructed void catalogues, 
allowing the measurements to retain their original void geometries.

The resulting stacked mass-ratio profiles are presented in
Figure~\ref{fig:void_mass_ratios}. In the asymptotic out-of-void regime
($r/R_{\rm v}\gtrsim1.5$), the three independently measured mass-ratio
estimators exhibit a high degree of consistency and satisfy the expected relation,
\begin{equation}
\mathcal{R}_{\rm g/dm}(r)
=
\mathcal{R}_{\rm g/sub}(r)
\times
\mathcal{R}_{\rm sub/dm}(r).
\end{equation}
Quantitatively, the profiles converge to approximately constant background
values of
$\mathcal{R}_{\rm g/dm}\approx0.0045$,
$\mathcal{R}_{\rm sub/dm}\approx0.30$,
and
$\mathcal{R}_{\rm g/sub}\approx0.015$.
The corresponding numerical consistency,
$0.015\times0.30=0.0045$,
provides an important validation of the independent multi-tracer measurements
on scales where environmental effects become negligible.

Moving towards smaller radii, both the galaxy-to-dark matter ratio,
$\mathcal{R}_{\rm g/dm}$, and the subhalo-to-dark matter ratio,
$\mathcal{R}_{\rm sub/dm}$, exhibit pronounced radial evolution. The ratios
are strongly suppressed within the innermost void regions
($r/R_{\rm v}\lesssim0.5$), increase steadily towards the void boundary,
and gradually approach their background values outside the void edge. This
behaviour indicates that both galaxies and massive subhaloes contribute a
smaller fraction of the total mass budget in underdense environments compared
with the surrounding regions. The effect is particularly pronounced for
$\mathcal{R}_{\rm sub/dm}$, which decreases from approximately $0.30$ in the
background region to $\sim0.05$ near the void centre.

These trends provide an empirical measurement of multi-tracer mass bias
inside cosmic voids. The systematic variation of
$\mathcal{R}_{\rm sub/dm}$ and $\mathcal{R}_{\rm g/dm}$ indicates that the
tracer--matter connection varies across different void environments.
The stronger radial variation of
$\mathcal{R}_{\rm sub/dm}$ suggests that massive subhaloes exhibit a more
pronounced environmental dependence than the galaxy population, although
both tracers show significant deviations from their background mass
fractions within void interiors.

Interestingly, despite the strong radial evolution observed in both
$\mathcal{R}_{\rm g/dm}$ and $\mathcal{R}_{\rm sub/dm}$, the mass ratio
$\mathcal{R}_{\rm g/sub}$ remains approximately constant over most of the
radial range. This indicates that, although galaxies and massive subhaloes
are both increasingly depleted towards void centres relative to the dark
matter field, their relative mass contribution is comparatively insensitive
to the void environment. These results suggest that the dominant
environmental dependence arises from the variation of the subhalo population
relative to the underlying matter distribution, while the galaxy--subhalo
mass connection remains comparatively stable across different void
environments. However, the large uncertainties in the deepest void regions
prevent a statistically significant assessment of possible environmental
dependence in $\mathcal{R}_{\rm g/sub}$.

Although the mean radial trends appear physically meaningful, the statistical
uncertainties reveal a more complicated picture. The SEM increases
substantially towards the inner void regions for both
$\mathcal{R}_{\rm g/dm}$ and $\mathcal{R}_{\rm g/sub}$, with the effect being
particularly severe for $\mathcal{R}_{\rm g/sub}$. The increased scatter
towards void interiors suggests that additional sources of variance are
present beyond the intrinsic void-to-void fluctuations.

Previous analyses by \citet{Zhang2026} demonstrated that the stacked radial
number-density profiles of SDSS galaxies and ELUCID subhaloes are highly
consistent when voids are identified using the same \texttt{REVOLVER}
framework. This result indicates that the two catalogues trace broadly
similar void environments on average. Nevertheless, the mass-ratio
measurements presented here exhibit substantially larger scatter towards void
interiors, suggesting that the geometric mismatch between independently
identified void structures may introduce additional variance.

A natural explanation is the residual geometric mismatch between the two void
catalogues. Because the independent-frame scheme measures galaxies, subhaloes,
and dark matter relative to their native void centres and effective radii,
offsets between matched SDSS and ELUCID voids can modify the enclosed mass
within the innermost radial bins. To quantify the impact of this effect, we
next examine the geometric offsets between matched void pairs and introduce an
alternative common-frame analysis.

\subsection{Coordinate offsets between matched voids}
\label{subsec:coordinate_offsets}

\begin{figure*}
\centering
\includegraphics[width=\textwidth]{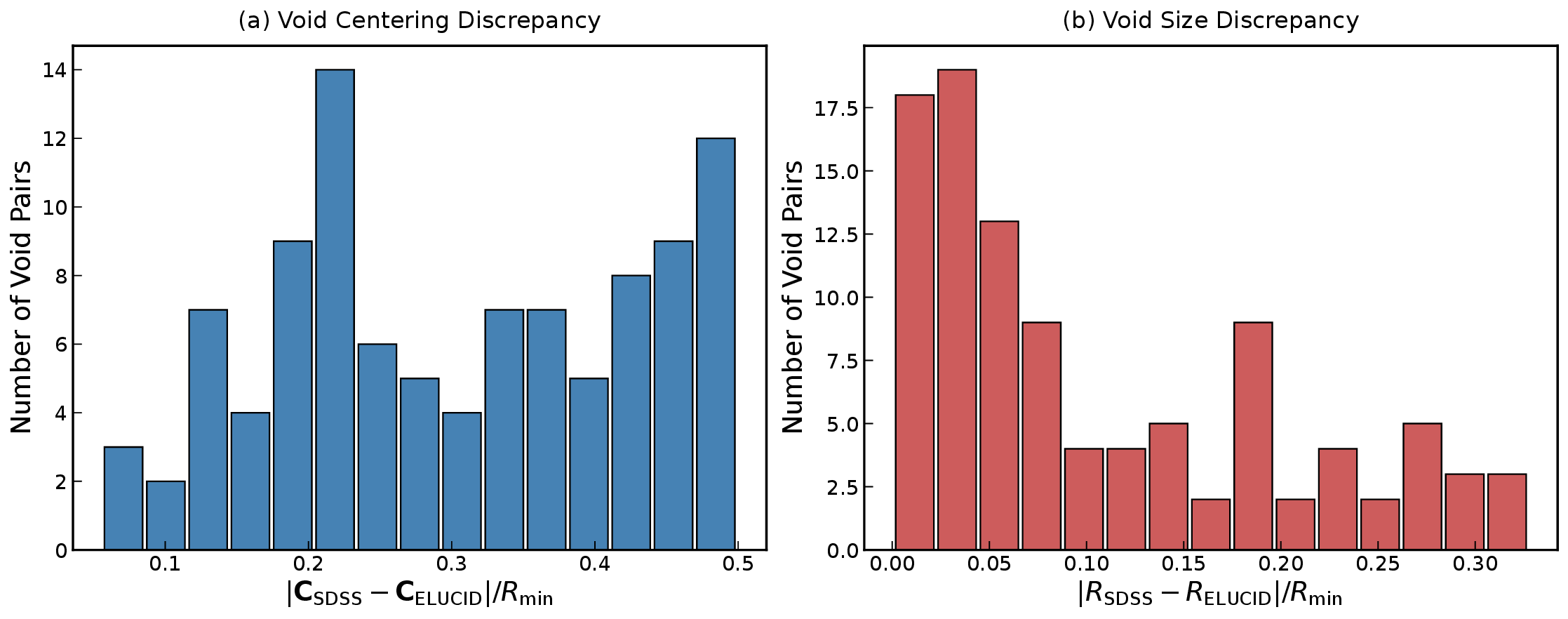}
\caption{
Distributions of the geometric differences for the 102 matched SDSS-ELUCID void pairs.
Panel (a) shows the normalised centre separation, $f_d = |\mathbf{C}_{\rm SDSS} - \mathbf{C}_{\rm ELUCID}|/R_{\rm min}$, 
where $R_{\rm min}=\min(R_{\rm SDSS},R_{\rm ELUCID})$.
Panel (b) shows the normalised radius difference, $f_{R}=|R_{\rm SDSS}-R_{\rm ELUCID}|/R_{\rm min}$.
}
\label{fig:void_offset_distribution}
\end{figure*}

The large scatter observed in the inner regions of the independent-frame
mass-ratio measurements motivates a more detailed examination of the
residual geometric differences between matched void pairs.
Although the SDSS and
ELUCID voids are selected as mutual best matches, their centres and
effective radii are not necessarily identical.

To quantify the precise structural discrepancies within our matched sample, we examine both the normalised centre offset
\begin{equation}
f_d =  \frac{|\mathbf{C}_{\rm SDSS} - \mathbf{C}_{\rm ELUCID}|}{R_{\rm min}}, 
\end{equation}
and the normalised radius variation
\begin{equation}
f_R = \frac {|R_{\rm SDSS} - R_{\rm ELUCID}|} {R_{\rm min}},
\end{equation}
where $R_{\rm min} = \min(R_{\rm SDSS}, R_{\rm ELUCID})$.

The distributions of these quantities for the 102 matched void pairs are
shown in Figure~\ref{fig:void_offset_distribution}.
As expected, the distribution of $f_d$ is truncated at
$f_d=0.5$ by construction, reflecting the proximity criterion
adopted during the void-matching procedure.
Nevertheless, the distribution extends over a broad range,
with many systems exhibiting
$f_d\gtrsim0.3$, indicating that non-negligible centre offsets
remain even among successfully matched systems. These offsets imply that
the galaxy and subhalo components entering the mass-ratio measurements
may be sampled from noticeably different physical regions when the
independent-frame scheme is adopted.

In contrast, the radius differences are generally modest. Although no
explicit constraint on the void-size difference is imposed during the
matching procedure, a large fraction of the matched pairs satisfy
$f_{R}<0.1$, and most systems exhibit
$f_{R}<0.2$. This result suggests that matched galaxy-defined and subhalo-defined
voids generally exhibit similar effective radii despite being identified
independently from different tracer populations.

The differing behaviour of the two distributions is noteworthy. While the
effective radii of matched voids generally agree well, their centre
positions can differ substantially. This indicates that coordinate
mismatches are primarily associated with void centering rather than void
size definitions.

Because the innermost void regions occupy only a small fraction of the total void volume, even modest centre offsets can substantially alter the tracer population enclosed within a given normalised radial shell. For example, a displacement of $f_d\sim0.3$ already corresponds to nearly one-third of the characteristic void radius, which can significantly
alter the inner profiles at $r/R_{\rm v}\lesssim0.5$ and artificially increase the statistical errors under the independent-frame scheme.

Conversely, the generally good agreement in void sizes
($f_R<0.1$ for the majority of pairs)
provides a physical justification for redefining a shared boundary. Since the characteristic scales are already well matched between the two populations, establishing a joint framework will not suffer from severe scale-mismatch distortions. 

Therefore, the measured geometric differences motivate us to mitigate
the centering offsets while preserving the characteristic void scale.
To achieve this, we next introduce a common-frame measurement scheme that places all tracer populations within a shared coordinate system based on the average positions and radii of the matched pairs.

\subsection{Common-frame measurements}
\label{subsec:common_frame}

\begin{figure*}
\centering
\includegraphics[width=\textwidth]{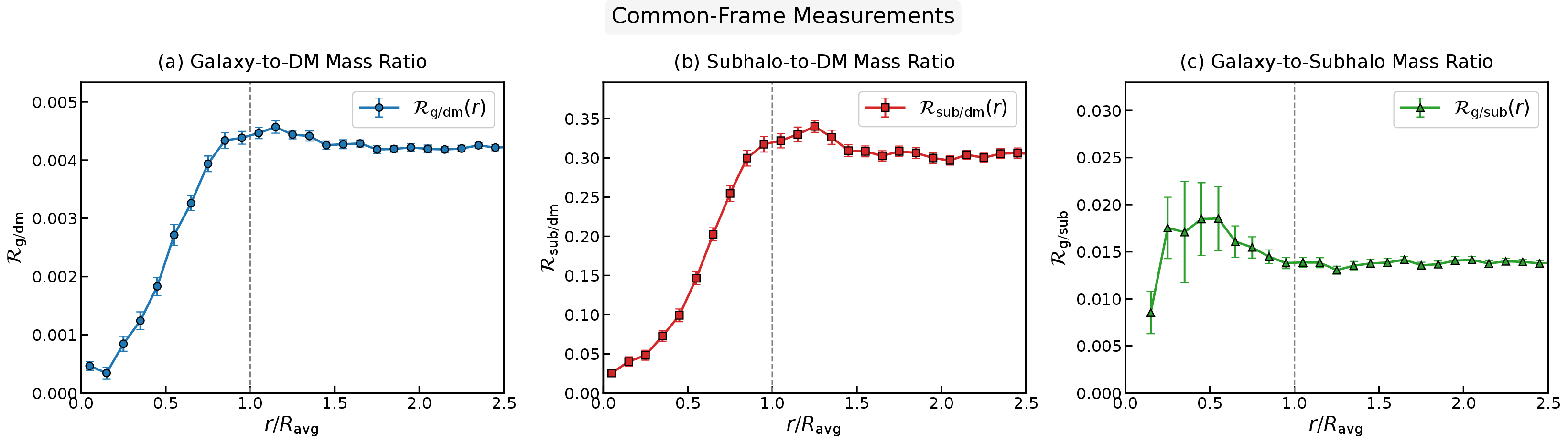}
\caption{
Radial profiles of the stacked mass ratios measured for the matched-void sample ($N_{\rm v} = 102$) using the common-frame scheme. Panels (a), (b), and (c) show the galaxy-to-dark matter ($\mathcal{R}_{\rm g/dm}$), subhalo-to-dark matter ($\mathcal{R}_{\rm sub/dm}$), and galaxy-to-subhalo ($\mathcal{R}_{\rm g/sub}$) mass ratios, respectively. The horizontal axis is scaled by the average void radius $R_{\rm avg}$, and the vertical dashed line marks the effective void boundary ($r/R_{\rm avg} = 1$). Error bars denote the standard error of the mean (SEM). Radial bins containing fewer than five valid void pairs are omitted.
}
\label{fig:void_mass_ratios_cfm}
\end{figure*}

For each matched void pair, we repeat the mass-ratio measurements using the common-frame scheme. Instead of adopting the native void definitions of the SDSS and ELUCID catalogues separately, all tracer populations are measured within a shared coordinate system defined by

\begin{equation}
\mathbf{C}_{\rm avg}
=\frac{\mathbf{C}_{\rm SDSS} + \mathbf{C}_{\rm ELUCID}}
{2},
\end{equation}
and
\begin{equation}
R_{\rm avg}
= \frac{R_{\rm SDSS}
+
R_{\rm ELUCID}}
{2}.
\end{equation}

Under this configuration, galaxies, subhaloes, and dark matter are sampled within the same geometric framework, thereby reducing the impact of centre and radius offsets between the two void catalogues.

Figure~\ref{fig:void_mass_ratios_cfm} presents the resulting mass-ratio profiles for the matched sample. Compared with the independent-frame measurements shown in Figure~\ref{fig:void_mass_ratios}, the overall shapes of the three mass-ratio profiles remain remarkably similar. In particular, both $\mathcal{R}_{\rm g/dm}$ and $\mathcal{R}_{\rm sub/dm}$ continue to increase from the void centre toward the void boundary before approaching approximately constant values at larger radii, indicating that the principal physical trends are robust against the choice of coordinate framework.

The most noticeable difference appears in the statistical uncertainties. For the galaxy-to-dark matter ratio, $\mathcal{R}_{\rm g/dm}$, the SEM is systematically reduced over nearly the entire radial range. The reduction is evident not only within the void interior but also beyond the void boundary. This behaviour suggests that a significant fraction of the scatter observed in the independent-frame measurement originates from coordinate offsets between the SDSS-defined and ELUCID-defined void centres. By forcing both tracer populations to be measured within a common spatial framework, these coordinate-induced fluctuations are substantially suppressed.

In contrast, the subhalo-to-dark matter ratio,
$\mathcal{R}_{\rm sub/dm}$, exhibits only minor differences between the
independent-frame and common-frame measurements, indicating that
coordinate offsets are not a dominant source of uncertainty for this
estimator.

The behaviour of the galaxy-to-subhalo ratio, $\mathcal{R}_{\rm g/sub}$, differs from that of $\mathcal{R}_{\rm g/dm}$. Although a moderate reduction in the SEM is visible outside the void boundary, the uncertainties remain large within the inner void regions ($r/R_{\rm avg}\lesssim0.6$). The common-frame scheme therefore reduces part of the coordinate-induced scatter but does not eliminate the substantial uncertainties associated with the innermost void shells.

The overall similarity of the mean profiles between the independent-frame and common-frame measurements indicates that the inferred mass-ratio trends are robust against the adopted coordinate definition. While the common-frame transformation generally reduces the statistical scatter, particularly for $\mathcal{R}_{\rm g/dm}$, it does not qualitatively alter the radial behaviour of any of the three mass-ratio estimators. The primary impact of the coordinate synchronization is therefore on the measurement uncertainty rather than on the underlying physical signal.

These results indicate that coordinate offsets are indeed an important source of uncertainty, particularly for measurements involving galaxies and dark matter. However, the persistence of large uncertainties in $\mathcal{R}_{\rm g/sub}$ demonstrates that coordinate offsets alone cannot fully account for the observed scatter. The remaining variance must therefore originate from limitations in the available tracer statistics rather than from coordinate offsets alone.

To identify the dominant source of this residual uncertainty, we next examine the effective number of valid void pairs contributing to each mass-ratio measurement as a function of radius.

\subsection{Tracer scarcity and the statistical limit of void-core measurements}
\label{subsec:tracer_scarcity}

\begin{figure}
\centering
\includegraphics[width=0.48\textwidth]{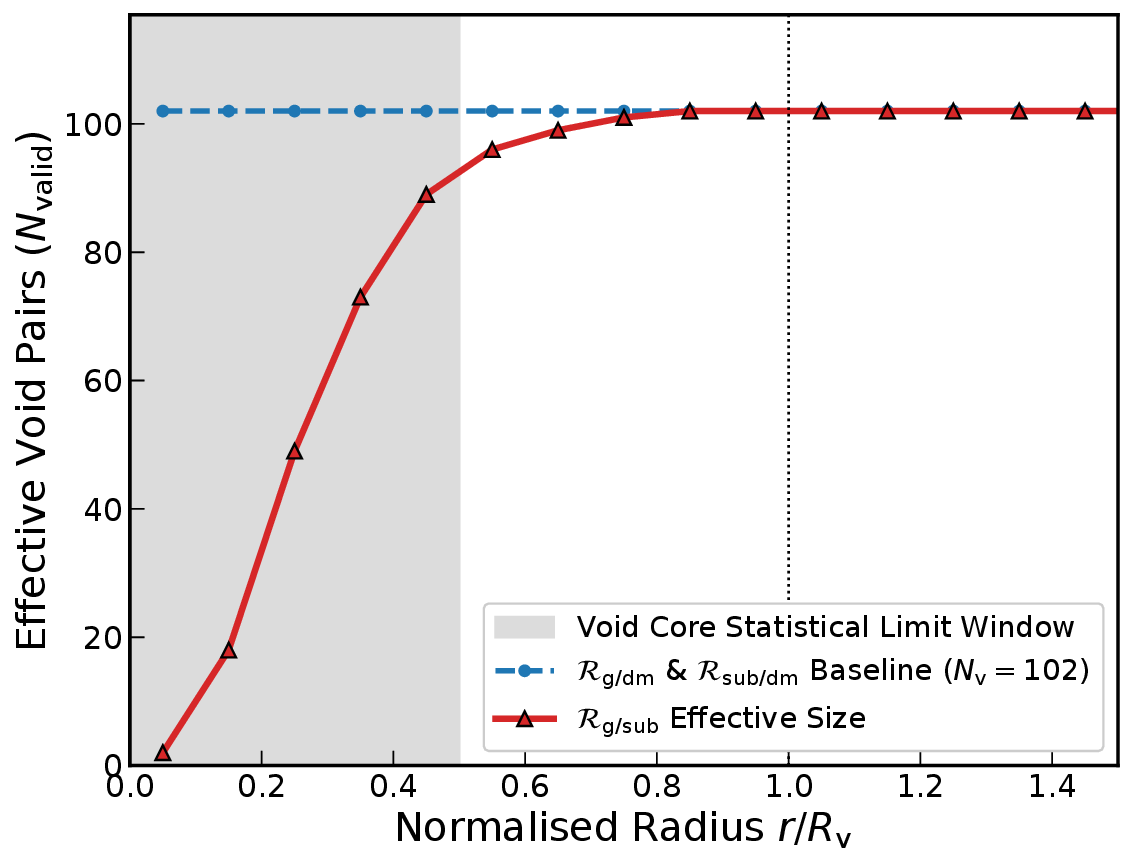}
\caption{
Effective number of valid void pairs,
$N_{\rm valid}$,
contributing to the mass-ratio measurements as a function of normalised radius.
The dashed curve corresponds to
$\mathcal{R}_{\rm g/dm}$
and
$\mathcal{R}_{\rm sub/dm}$,
while the solid curve shows
$\mathcal{R}_{\rm g/sub}$.
The shaded region highlights the
\emph{Void Core Statistical Limit Window},
where the effective sample size decreases rapidly and the measurements become increasingly dominated by small-number statistics.
}
\label{fig:effective_size}
\end{figure}

The effective number of valid void pairs provides a direct diagnostic of
the statistical power available for estimating the uncertainty of each
mass-ratio measurement. For a given radial bin, a void pair is
considered valid only if the denominator of the corresponding mass ratio
is non-zero.

Figure~\ref{fig:effective_size} shows the radial dependence of
$N_{\rm valid}$ for the three mass-ratio statistics measured under the
independent-frame scheme.
For both
$\mathcal{R}_{\rm g/dm}$
and
$\mathcal{R}_{\rm sub/dm}$,
the effective sample size remains equal to the full matched sample
($N_{\rm valid}=102$)
across the entire radial range.
This behaviour reflects the fact that dark matter is represented by a
continuous particle field, ensuring non-zero denominator masses in every
radial shell.

In contrast,
$\mathcal{R}_{\rm g/sub}$
shows a rapid decline in effective sample size toward the void centre.
Although nearly all matched void pairs contribute close to and beyond the
void boundary, the number of valid systems decreases sharply within the
void interior.
At
$r/R_{\rm v}=0.25$,
fewer than half of the matched void pairs remain available for the
measurement (49 out of 102), while at
$r/R_{\rm v}=0.05$
the effective sample size collapses to only two systems.

This reduction originates from the extreme scarcity of massive
subhaloes
($\log_{10}(M_{\rm sub}/h^{-1}M_\odot)\ge11.8$)
within void interiors.
For many matched void pairs, no subhalo satisfying this mass threshold
is present in the innermost radial shells, resulting in vanishing
denominator masses and therefore undefined values of
$\mathcal{R}_{\rm g/sub}$.
Consequently, only a small fraction of the matched sample contributes to
the measurement at the smallest radii.

The strong decline of
$N_{\rm valid}$
provides a natural explanation for the behaviour of
$\mathcal{R}_{\rm g/sub}$
observed in both the independent-frame and common-frame measurements.
Even after coordinate offsets between matched voids are mitigated, the
statistical power of the measurement remains fundamentally limited by the
small number of available systems within the void cores.
As a result, the innermost regions become increasingly dominated by
small-number statistics and exhibit substantially larger uncertainties
than those associated with
$\mathcal{R}_{\rm g/dm}$
or
$\mathcal{R}_{\rm sub/dm}$.

These results demonstrate that coordinate offsets and tracer scarcity
play fundamentally different roles in multi-tracer void analyses.
Coordinate offsets constitute an important source of uncertainty when
comparing tracer populations defined in different void catalogues, as
demonstrated by the improvement of
$\mathcal{R}_{\rm g/dm}$
under the common-frame scheme.
In contrast, for measurements involving rare discrete tracers such as
massive subhaloes, the dominant limitation within void interiors arises
from the rapid collapse of the effective sample size itself.
Consequently, mass-ratio measurements within the
\emph{Void Core Statistical Limit Window}
($r/R_{\rm v}\lesssim0.5$)
should be interpreted with particular caution, regardless of the adopted
coordinate framework.

\section{Discussion}\label{sec:discussion}

\subsection{Environmental dependence of tracer--matter connections}

The radial behaviour of the three mass-ratio estimators provides
additional insight into how different tracers respond to the void
environment. While both $\mathcal{R}_{\rm g/dm}$ and
$\mathcal{R}_{\rm sub/dm}$ decrease significantly towards void centres,
the nearly constant profile of $\mathcal{R}_{\rm g/sub}$ indicates that
the relative galaxy-to-subhalo mass contribution shows only a weak
dependence on the large-scale void environment.

This result suggests that the dominant environmental modulation of void
mass bias occurs primarily through the subhalo population and its
contribution relative to the underlying matter distribution, rather than
through a strong variation in the relative galaxy-to-subhalo connection.
The relatively stable galaxy-to-subhalo mass ratio suggests that
the environmental variation in the galaxy-to-dark matter mass ratio is
largely associated with the changing subhalo-to-dark matter relation,
rather than with a comparably strong change in the relative mass
contribution of galaxies and subhaloes.

This interpretation implies that the environmental dependence of tracer
mass bias in voids is closely related to the abundance and spatial
distribution of massive subhaloes, while the relative contribution of
galaxies with respect to their host subhaloes remains comparatively
stable. Our results therefore point to the subhalo population as
an important component of the environmental dependence of tracer--matter
mass bias, while the galaxy-to-subhalo mass contribution appears less
sensitive to void-centric environment over the range probed here.
Such a picture is consistent with previous findings that the
galaxy--halo connection is closely linked to intrinsic halo properties
and assembly histories \citep{Yang2018, Zhang2021b, Xu2024}. At the same
time, recent studies have demonstrated that underdense environments can
also influence galaxy properties, including stellar populations and star
formation activity \citep{Song2026b}. A complete understanding of the
environmental dependence of the galaxy--subhalo connection therefore
requires additional galaxy properties, such as stellar mass, colour, and
star formation activity, which are beyond the scope of the present work
and will be investigated in future studies.

\subsection{Implications of geometric differences for multi-tracer measurements}

A central result of this work is that voids identified from different
tracer populations, even when tracing the same underlying large-scale
underdensity, can exhibit non-negligible geometric differences.
Although the matched SDSS and ELUCID voids generally occupy the same
cosmic environments and exhibit similar effective radii, differences
remain in their centre positions and void boundaries.

Our common-frame analysis demonstrates that these geometric differences
can introduce non-negligible uncertainties into multi-tracer
measurements. In particular, the systematic reduction of the SEM in the
galaxy-to-dark matter ratio after adopting a common coordinate system
indicates that a substantial fraction of the scatter observed in
independent-frame measurements originates from geometric differences
between the two void catalogues rather than from intrinsic variations in
the tracer--matter relation. This effect becomes especially important
within void interiors, where the sampled volumes are small and modest
coordinate offsets can significantly alter the enclosed tracer
populations.

Although this study focuses on matched voids constructed from SDSS
galaxies and the ELUCID constrained simulation, the underlying issue is
much more general. Similar geometric differences are expected whenever
voids are identified from different tracer populations, even within the
same simulation volume, such as galaxies, subhaloes, haloes, and dark
matter particles. Consequently, caution is required when interpreting
differences between void profiles derived from distinct catalogues. The
common-frame approach adopted here provides a simple and computationally
inexpensive way to mitigate such coordinate-induced scatter.

\subsection{Statistical limits of void-core measurements}

Despite the improvement achieved by the common-frame scheme, the
galaxy-to-subhalo mass ratio remains highly uncertain within the inner
regions of voids. Our effective-sample-size analysis demonstrates that
this limitation is not primarily caused by coordinate offsets, but by
the limited abundance of massive subhaloes within the matched void
sample.

For the galaxy-to-dark matter and subhalo-to-dark matter ratios, the
effective sample size remains equal to the full matched sample across
all radii because dark matter is continuously distributed throughout the
simulation volume. In contrast, the effective sample size associated
with the galaxy-to-subhalo ratio decreases rapidly toward the void
centre, because an increasing fraction of matched voids contain no
subhaloes above the adopted mass threshold within the innermost radial
shells. Consequently, the number of valid systems contributing to the
measurement becomes severely reduced within the void-core region
($r/R_{\rm v}\lesssim0.5$), causing the uncertainty to be dominated by
small-number statistics.

This result highlights an important practical limitation of
observationally anchored void studies. Although constrained
simulations provide access to the dark matter and halo information that
is unavailable in observations, the number of matched void systems is
still limited by the available survey volume and by the abundance of
the tracer population under consideration. In particular, measurements
involving rare tracer populations, such as
massive subhaloes, remain challenging in the deepest underdense regions
when based on current survey volumes.

Future surveys such as DESI \citep{Yang2021}, combined with
next-generation constrained simulations \citep{Hong2026}, will
substantially increase the number of available matched void systems and
improve tracer statistics within void interiors. Such datasets
will provide a promising route toward more precise multi-tracer
measurements.

\subsection{Toward observationally anchored measurements of void mass bias}

The principal contribution of this work is the development and
demonstration of an observationally anchored framework for comparing
the mass contributions of multiple tracers within corresponding
underdense environments. Rather than replacing either observational
measurements or simulation-based studies, this framework combines the
observational connection provided by the SDSS galaxy distribution with
the subhalo and dark matter information available from the constrained
ELUCID simulation.

Most previous studies of tracer--matter relations in cosmic voids have
relied either on observational galaxy catalogues, which provide access
to luminous tracers but not directly to the underlying dark matter
distribution, or on simulations, which provide complete information
about dark matter and halo populations but are not generally tied to
the specific large-scale structures observed in the real Universe.
The matched SDSS--ELUCID framework adopted here bridges these
two aspects by associating observed galaxy-defined voids with
counterpart voids in an observationally constrained realisation of the
local large-scale structure. This allows the relative mass
contributions of galaxies, subhaloes, and dark matter to be examined
within corresponding cosmic environments, while retaining the
observational origin of the galaxy sample
\citep{WangHuiyuan2016,Zhang2025,Zhang2026}.

The value of this observational anchoring is therefore not
simply the availability of additional simulated quantities, but the
ability to study them in environments that are constrained by the
observed local galaxy distribution. At the same time, the matched
framework provides a controlled setting in which methodological effects
can be examined alongside the physical measurements. In the present
analysis, for example, comparisons between independent and common
coordinate frameworks allow the impact of void-centre offsets to be
assessed, while the effective-sample-size analysis reveals how tracer
abundance limits the statistical precision of mass-ratio measurements
in the innermost void regions. These aspects would be difficult to
separate using either observational catalogues or unconstrained
simulations alone.

Although the current analysis is limited by the modest number of
matched voids available in the SDSS volume, it demonstrates the
feasibility of empirical multi-tracer mass-bias measurements
within observationally constrained environments. 
With larger survey volumes and improved tracer statistics, this framework can be
applied to increasingly detailed multi-tracer comparisons within
corresponding void environments.

\section{Conclusions}
\label{sec:conclusion}

In this work, we combined the SDSS galaxy catalogue with the ELUCID
constrained simulation to perform observationally anchored
multi-tracer measurements of galaxies, subhaloes, and dark matter
within matched cosmic voids. By independently identifying voids in the
SDSS galaxy and ELUCID subhalo catalogues and subsequently constructing
a one-to-one matched sample, we obtained 102 matched void pairs
that occupy corresponding large-scale underdense environments
in the local Universe.

Using these matched systems, we measured the radial profiles of three
mass-ratio statistics,
$\mathcal{R}_{\rm g/dm}$,
$\mathcal{R}_{\rm sub/dm}$,
and
$\mathcal{R}_{\rm g/sub}$.
Our main conclusions are summarized as follows.

\begin{enumerate}

\item
Both the galaxy-to-dark matter ratio and the subhalo-to-dark matter
ratio exhibit pronounced radial dependence within cosmic voids. The
ratios decrease toward void centres and increase toward the void
boundary, indicating that galaxies and massive subhaloes become
progressively more depleted than the underlying dark matter field in the
deepest underdense regions.

\item
The galaxy-to-subhalo ratio remains approximately constant over most of
the radial range outside the innermost void regions, suggesting that the
galaxy--subhalo connection is comparatively insensitive to the
large-scale void environment. The observed evolution of the
galaxy-to-dark matter ratio therefore appears to arise primarily from
the environmental dependence of the subhalo population rather than from
a substantial variation in the galaxy--subhalo connection itself.

\item
Matched SDSS and ELUCID voids exhibit generally consistent effective
radii but non-negligible centre offsets. Adopting the common-frame
measurement scheme significantly reduces the statistical uncertainties
of the galaxy-to-dark matter measurements, demonstrating that part of
the scatter observed in the independent-frame analysis originates from
coordinate offsets between independently identified void catalogues.

\item
The large uncertainties associated with the galaxy-to-subhalo ratio
persist even after coordinate synchronization. By examining the radial
distribution of the effective number of contributing void systems, we
show that this limitation is primarily caused by the severe scarcity of
massive subhaloes within void interiors. In the innermost radial bins,
only a small fraction of the matched void sample contains any massive
subhalo, leading to a rapid reduction of the effective sample size and
fundamentally limiting the statistical precision attainable within void
cores.

\end{enumerate}

These results demonstrate both the scientific potential and the
practical limitations of observationally anchored multi-tracer
measurements in cosmic voids. Constrained simulations such as ELUCID
provide access to the dark matter and halo information unavailable from
observations, while preserving a connection to the observed large-scale
environment. However, the extreme sparsity of massive
tracers in void interiors imposes a fundamental statistical limitation
on the precision of certain multi-tracer measurements.

The main methodological value of this approach is that it
combines these complementary sources of information within corresponding
void environments. Observed galaxy-defined voids retain the connection
to the real local Universe, while their counterparts in the constrained
simulation provide access to subhalo and dark matter information that
cannot be obtained directly from the galaxy catalogue. This makes it
possible to investigate relative tracer mass contributions while also
quantifying the effects of void geometry, coordinate consistency, and
tracer abundance on the measurements.

The methodology developed in this work is particularly relevant for the
forthcoming ELUCID--DESI project \citep{Hong2026}. By combining the
substantially larger survey volume of DESI with next-generation
constrained simulations, future studies will increase the available void
sample by more than an order of magnitude and significantly improve
tracer statistics within void interiors. Such datasets will substantially 
improve the statistical precision of multi-tracer mass-bias measurements, 
particularly in the sparsely sampled inner regions of cosmic voids.

More broadly, the present work demonstrates the feasibility of
using observationally anchored, matched void environments as a
complementary setting for multi-tracer studies of the cosmic web. The
framework provides a basis for comparing different tracer populations
within corresponding large-scale environments while retaining the
observational connection to the local Universe.

\section*{Acknowledgements}

This research is funded by various grants, including the National Natural Science Foundation of China (Nos. 12273088, 12633001, 12595313, 12103037), the National SKA Program of China (grant No. 2025SKA0150100), and the National Key R\&D Programme of China (2023YFA1607800, 2023YFA1607804). Additional support comes from the CSST project (Nos. CMS-CSST-2021-A02, CMS-CSST-2025-A04), the CAS Project for Young Scientists in Basic Research (No. YSBR-092), Fundamental Research Funds for Central Universities, the 111 project (No. B20019) and the Shanghai Natural Science Foundation (grant No.19ZR1466800, 23JC1410200, ZJ20223-ZD-003). PW acknowledge financial support by the NSFC (No. 12473009), and also sponsored by Shanghai Rising-Star Program (No.24QA2711100). 

This work is also supported by the High-Performance Computing Resource in the Core Facility for Advanced Research Computing at Shanghai Astronomical Observatory.

This work used {\tt Astropy:} a community-developed core Python package
and an ecosystem of tools and resources for astronomy \citep{Astropy2013, 
Astropy2018, Astropy2022}. 

Funding for the Sloan Digital Sky Survey IV has been provided by the
Alfred P. Sloan Foundation, the U.S. Department of Energy Office of
Science, and the Participating Institutions. SDSS acknowledges support
and resources from the Centre for High-Performance Computing at the
University of Utah. The SDSS website is www.sdss.org.

SDSS is managed by the Astrophysical Research Consortium for the
Participating Institutions of the SDSS Collaboration including the
Brazilian Participation Group, the Carnegie Institution for Science,
Carnegie Mellon University, the Chilean Participation Group, the
French Participation Group, Harvard-Smithsonian Center for
Astrophysics, Instituto de Astrof{\'i}sica de Canarias, The Johns
Hopkins University, Kavli Institute for the Physics and Mathematics of
the Universe (IPMU)/University of Tokyo, Lawrence Berkeley National
Laboratory, Leibniz Institut f{\"u}r Astrophysik Potsdam (AIP),
Max-Planck-Institut f{\"u}r Astronomie (MPIA Heidelberg),
Max-Planck-Institut f{\"u}r Astrophysik (MPA Garching),
Max-Planck-Institut f{\"u}r Extraterrestrische Physik (MPE), National
Astronomical Observatories of China, New Mexico State University, New
York University, University of Notre Dame, Observat{\'o}rio Nacional/
MCTI, The Ohio State University, Pennsylvania State University,
Shanghai Astronomical Observatory, United Kingdom Participation Group,
Universidad Nacional Aut{\'o}noma de M{\'e}xico, University of
Arizona, University of Colorado Boulder, University of Oxford,
University of Portsmouth, University of Utah, University of Virginia,
University of Washington, University of Wisconsin, Vanderbilt
University, and Yale University.

\section*{Data availability}
The data underlying this article will be shared on reasonable request with
the corresponding author.

\bibliographystyle{mnras}
\bibliography{bibliography}

% Don't change these lines
\bsp    % typesetting comment
\label{lastpage}
\end{document}